# Quantum well infrared photodetectors hardiness to the non ideality of the energy band profile

Emmanuel Lhuillier, Nicolas Péré-Laperne, Isabelle Ribet-Mohamed, Emmanuel Rosencher
*ONERA, centre de Palaiseau, Chemin de la Hunière - FR 91761 Palaiseau cedex, France.*

Gilles Patriarche
*Laboratoire de Photonique et de Nanostructures, LPN/UPR20 - CNRS Route de Nozay, 91460 Marcoussis, France.*

Amandine Buffaz, Vincent Berger
*Matériaux et Phénomènes quantiques, Case 7021, université Denis Diderot - Paris 7, Bâtiment Condorcet, 75205 Paris cedex 13, France.*

Alexandru Nedelcu, Mathieu Carras
*Alcatel-Thales III-V Lab, Campus de Polytechnique, 1 Avenue A. Fresnel, 91761 Palaiseau cedex, France.*

We report results on the effect of a non-sharp and disordered potential in Quantum Well Infrared Photodetectors (QWIP). Scanning electronic transmission microscopy is used to measure the alloy profile of the structure which is shown to present a gradient of composition along the growth axis. Those measurements are used as inputs to quantify the effect on the detector performance (peak wavelength, spectral broadening and dark current). The influence of the random positioning of the doping is also studied. Finally we demonstrate that QWIP properties are quite robust with regard to the non ideality of the energy band profile.

## I. Introduction

Optoelectronic devices properties result of the ability of the growth process to realize sharp interfaces giving rise to an energy band profile (EBP) which is close to the initial design. Thanks to the progress of the growth methods (Molecular Beam Epitaxy and Molecular Organic Vapour Phase Epitaxy) during the last decade, it is possible to obtain structures very close to the nominal design with a very low defect density. Nonetheless residual non ideality of the growth process may drastically decrease the final performances.[1]

In this paper we address the issue of the hardiness of high wavelength (λ~15µm) quantum well infrared photodetectors due to the non ideality of the EBP. Therefore, transmission electronic microscopy is used to point out the deviation between the grown structure (no post growth treatment such as thermal annealing[2]) and the nominal structure. Then, numerical simulations are performed to evaluate the effect of the deviation on the EBP. Finally, the simulated EBP allows us to calculate the effects on the infrared detector performances (peak wavelength, spectral broadening, dark current). Those effects are studied on a Quantum Well Infrared Photodetector (QWIP) structure operating in the tunnel regime, *i.e.* at low temperature. Indeed the tunnel regime is very interesting for our study since the transport is very sensitive to the EBP, contrary to the thermionic regime.

Two different sources of non ideality are considered in this paper: the non-sharpness of the profile and the presence of disorder. (i) Concerning the first one, electronic microscopy measurements[3] give a direct access to the interface profile. We predict that this non-sharpness of the EBP leads to a redshift of the spectral response and to an increase of the dark current. (ii) The second source of non ideality, the disorder induced by the presence of ionized charges, leads to a blue shift of the transition and to a reduced dark current. Finally, we underline the great role of the final subband on the device properties and show that QWIP is quite robust despite the non ideality of the EBP.

## II. Experimental and modelling considerations

### A. Sample of interest

The studied structure is a forty periods $GaAs/Al_{0.15}Ga_{0.85}As$ QWIP.[4] The barrier width is 35nm, and the well width is 7.3nm. Each well is Si-doped in the central third of the well with a nominal sheet concentration of $3\times10^{11}cm^{-2}$. The responsivity of the structure is peaked at 14.5µm with a full width at half maximum (FWHM) of 2µm (equivalent to 12meV), (inset of FIG. 1). Measurements were performed at very low temperature in order to reduce the dark current, since the detector is dedicated to very low infrared photon flux. A cryogenic set-up was used for these measurements from 4K to 25K. Under these conditions, the dark current originates from the tunnel regime. Electrons are driven by the hopping between ground states of two adjacent wells. The expected bias operating point is around -1V, thus the structure will be biased with an electric field close to this value in the following.

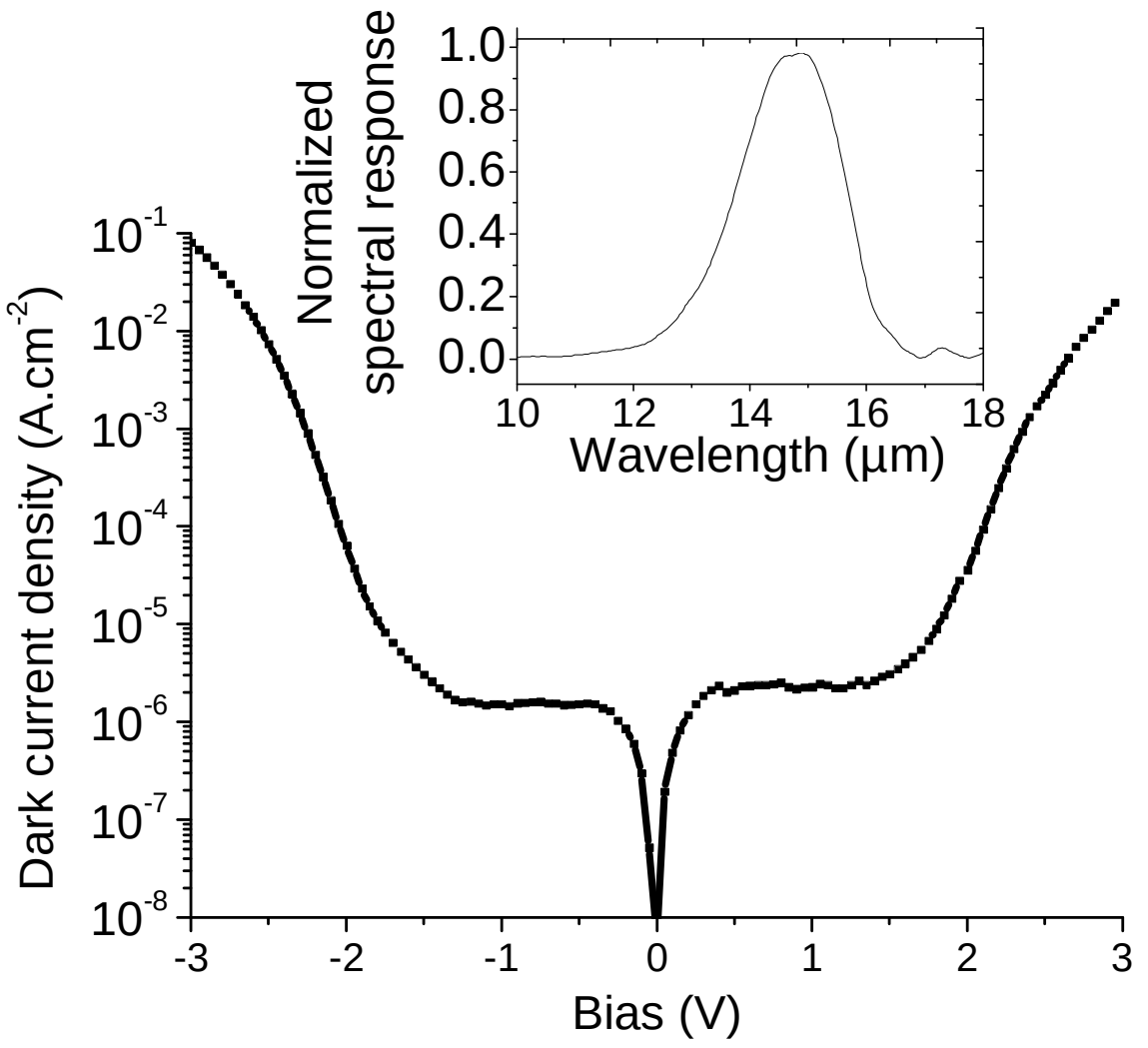

FIG. 1 Dark current as a function of the voltage bias at T=4K. Inset: normalized spectral response at T=4K and V=-1V.

## B. Electronic states and scattering simulation

Energies and waves functions are obtained using a Schrödinger code based on a shooting method.[5] Our scattering simulation, based on a hopping model between two dimensional states, allows us to evaluate both the intrawell and interwell scattering times. The following scattering processes are taken into account in our simulation: interaction between electron and LO phonons (LO), acoustical phonons (AC), alloy disorder (AL), interface roughness (IR), ionized impurities (II). Details of the scattering rates calculation are given in ref 6. The interaction between carriers has been neglected since this interaction is not dominant in this device[6]. Simulations are performed in a double quantum well structure.

# III. Interface non-sharpness

Usually the modelling of the optoelectronic properties in heterostructures is performed assuming the sharpness of the GaAs-AlGaAs interfaces. Thanks to its high resolution and great Z contrast, scanning transmission electron microscopy (STEM) operating in high angle annular dark-field imaging mode (HAADF) is a good probe of the alloy profile. STEM highlights the non-sharpness of the interfaces. Parameters extracted from those measurements are used to quantify the effect of the non-sharpness on the optical and transport properties of the QWIP device.

## A. STEM measurements

STEM measurements have been realized on a 9µm quantum cascade detector[7,8] (QCD) sample, see FIG. 2. This is a forty periods structure (each is 44nm long, composed of seven wells in each). The growth parameters are described in ref 7. The growth of this sample is expected to be similar enough to the QWIP growth (same MBE system, similar material, same growth temperature), to use the results obtained from the QCD on the QWIP device.

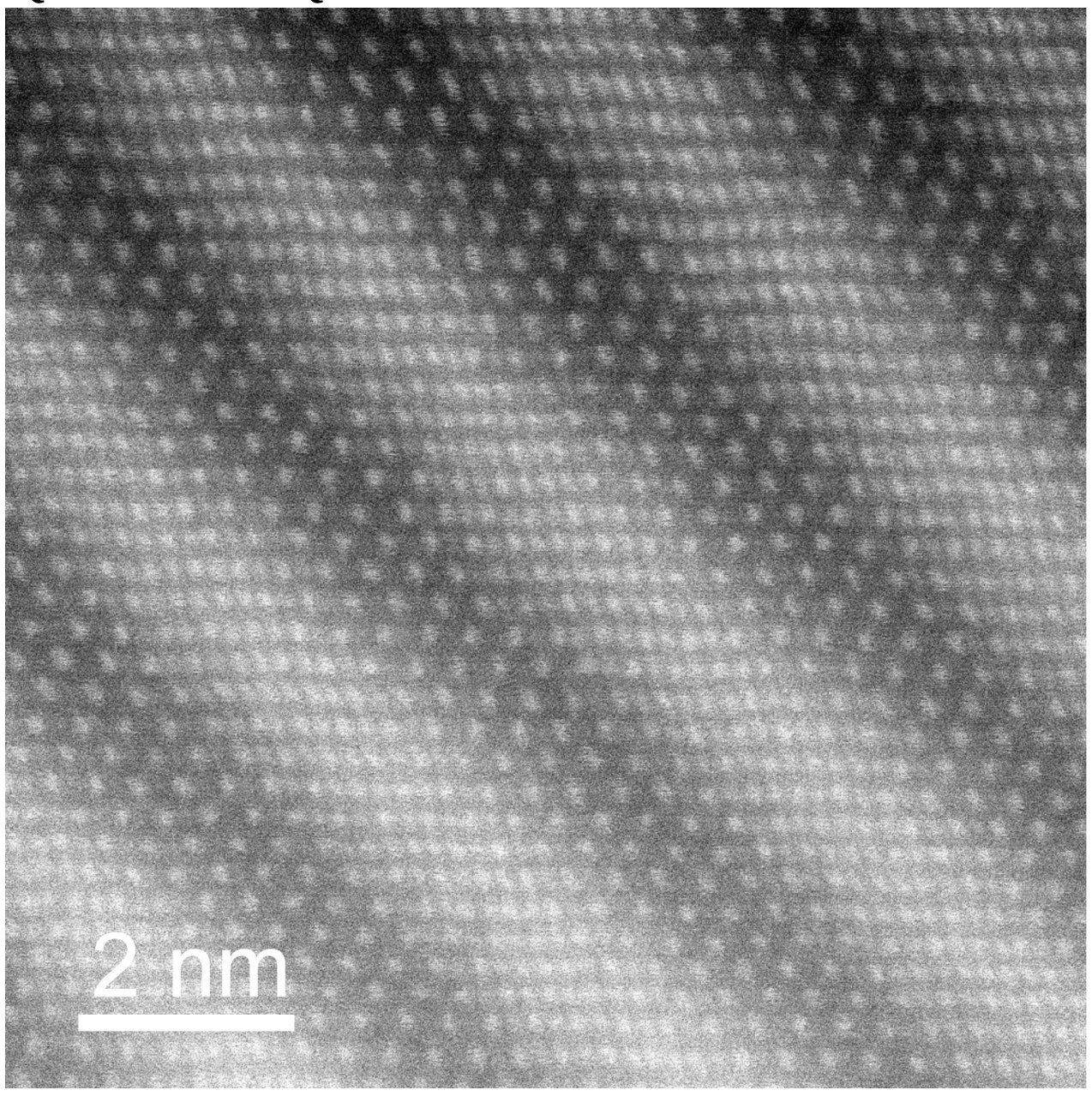

FIG. 2 High resolution Z-contrast image of the 9µm QCD structure prepared by <100> cross-section.

STEM images were acquired at 200 keV with a JEOL JEM 2200FS scanning transmission electron microscope equipped with a CEOS aberration-corrector. HAADF-STEM images (also called high-resolution Z contrast images) were obtained with a half-angle probe of 30 mrad, the inner and outer half-angles of the annular detector (called upper-HAADF detector in this machine) were, respectively, 100 mrad and 170mrad. The resolution of the microscope is expected to be 1Å. Two series of images were respectively performed in the [110] and [100] directions. FIG. 3(b) presents the alloy profile in the growth direction and reveals a gradient of composition at the interface. From the intensity profile of the high resolution image, the gradient extension is evaluated to range from two to four monolayers. Such a value is obtained after averaging over several interfaces.

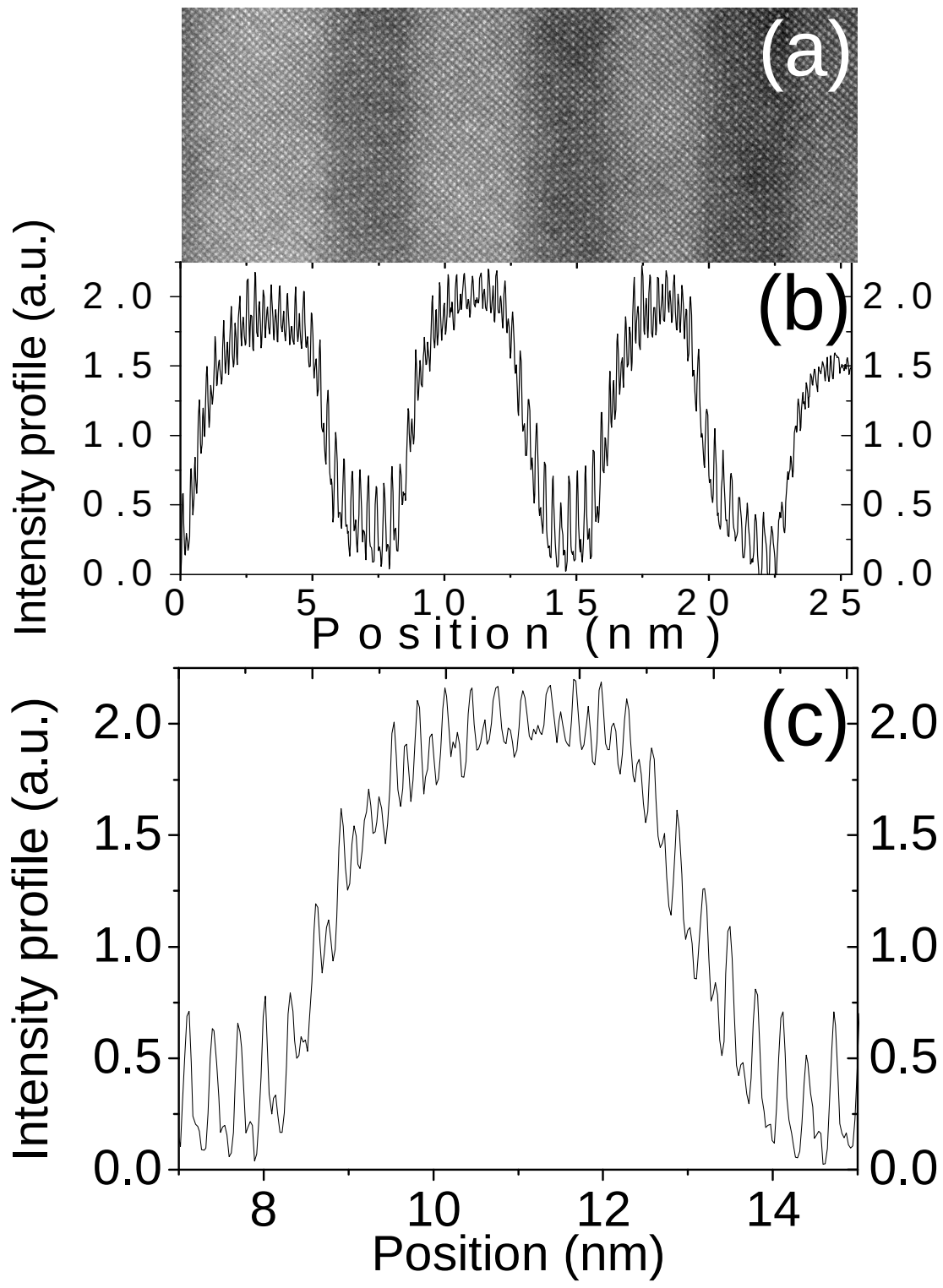

FIG. 3 (a) High resolution Z-contrast image of the QCD structure prepared by <100> cross-section; (b) Corrected intensity profile along the growth direction. (c) Zoom of the intensity profile over two interfaces.

## B. Origin of the non-sharpness

The identification and understanding of non-sharp interfaces is crucial for the modelling of the characteristics and for the reduction of this non ideality.

Due to the moderate growth temperature of the GaAs (500-600°C range), non-sharpness cannot come from a bulk diffusion of the III type element. Actually the scattering of the species is effectively possible in GaAs devices but for higher growth temperature[9] (above 800°C). This effect has already been used to tune the detected wavelength and to broaden the transition of QWIPs.[10]

Segregation can be mentioned as a possible source of non-sharpness. This effect is non negligible in GaAs/InGaAs devices.[11] The segregation of the doping species has also been mentioned in GaAs/AlGaAs structures and generally leads to an asymmetry in the I(V) curve, as mentioned in ref 12. As the alloy profile presents a very low dissymmetry, segregation is believed to play a minor role.

Interface non-sharpness may mostly result from the opening and closing time of the aluminum shutter. (i) First, the opening of the shutter leads to a decrease of the aluminum cell temperature which decreases the associated flux. Arthur[13] evaluated the decay time linked to this process to 30s, whereas the growth rate is around 2Å.s$^{-1}$ making this process compatible with the observed alloy gradient extension. (ii) Secondly, during the opening/closing process the aluminum cell is partly hidden, which results in a limited flux of Aluminum. The shutter opening/closing time is typically in the 100-200ms range[14], thus this effect will only act on one monolayer (ML). We believe that those two combined effects, due to the shutter opening and closing time, lead to this composition gradient of a few MLs. As a consequence, for two different samples grown in the same MBE system, a similar non-sharpness is expected, which *a posteriori* validates our choice to study a QCD sample and to extrapolate the result to a QWIP.

## C. Modelling of the energy band profile

In the following, we choose to model the non ideal EBP introducing interface non-sharpness. The potential profile for each well is expected to follow an error function shape, see the inset of FIG. 4:

$$V(z) = V_b \left( 1 - 1/2 \sum_{well} erf\left(\frac{z - Z_r}{L_d}\right) - erf\left(\frac{z - Z_l}{L_d}\right) \right)$$

Where $L_d$ is the gradient extension and $Z_r$ ($Z_l$) is the position of the right (left) interface of the well. The mass profile is similarly affected.

**D. Influence of the smoothing on the optoelectronic properties**

Using the previous profile we can compute the associated wave function and energies.

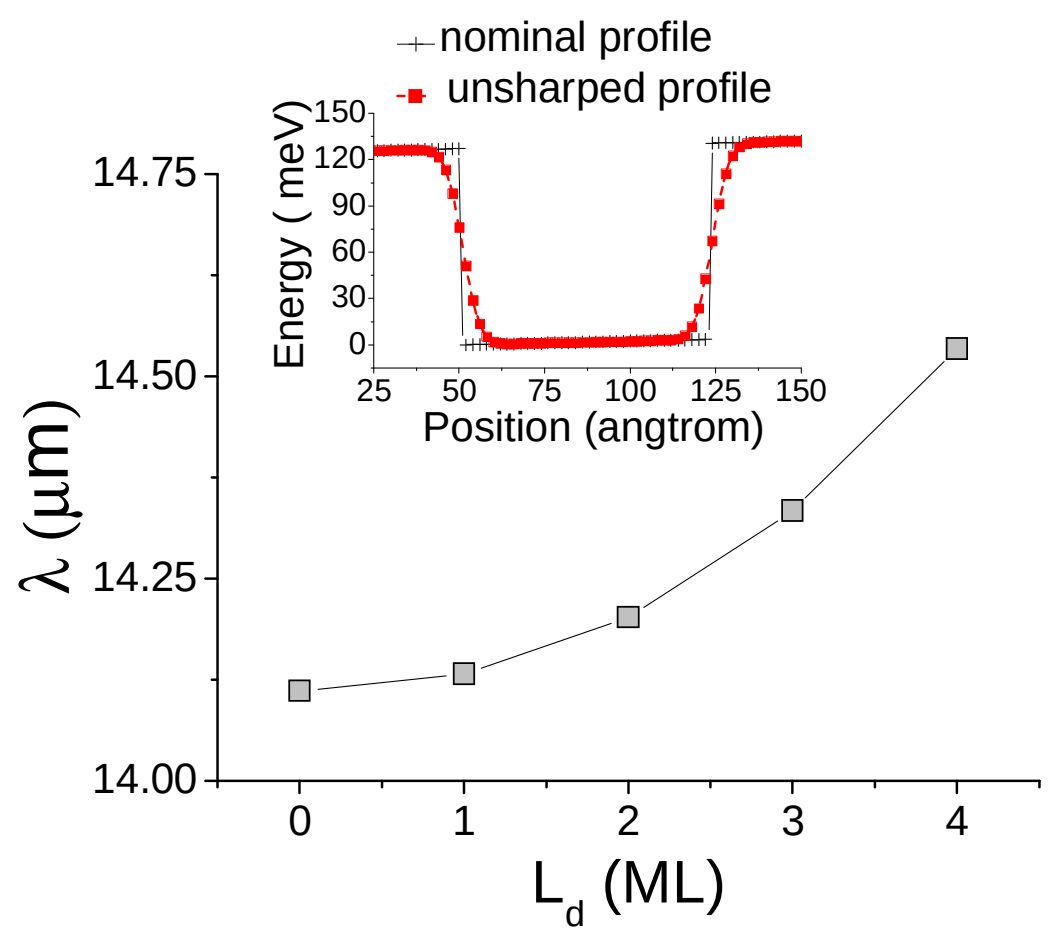

FIG. 4 Evolution of the wavelength associated to the transition between the ground state and excited state as a function of gradient extension in monolayers (ML). Inset: EBP from a single well for $L_d$=2ML and F=5kV.cm$^{-1}$.

FIG. 4 presents the influence of the alloy gradient extension on the energy transition between the ground state and the first excited state. As expected, a redshift[15] occurs on this transition resulting from a reduced confinement of the electrons into the well. STEM measurements have exhibited a fluctuation of the gradient extension typically one monolayers for a three monolayers value of the gradient extension, which results in a 0.15µm (0.9meV) heterogeneous broadening of the transition.

tab. I Magnitude of the interwell scattering rate with and without disorder, for T=10K, F=5kV.cm$^{-1}$. Values are evaluated for a three monolayers gradient extension.

| Interaction | Interwell scattering rate without disorder (Hz) | Interwell scattering rate with disorder (Hz) |
|---|---|---|
| LO | $1.2\times10^{-18}$ | $4.3\times10^{-19} \pm 9.6\times10^{-20}$ |
| AC | $2.8\times10^{-2}$ | $1.2\times10^{-2} \pm 2.8\times10^{-3}$ |
| AL | 1.7 | 0.59 ± 0.15 |
| IR | 2.6 | 0.9 ± 0.23 |
| II | 6.4 | 2.4 ± 0.5 |

The interwell scattering rate exhibits a monotonic increasing behaviour with the gradient extension, see FIG. 5, a consequence of a lower confinement of the first bound state. Assuming an inversely proportional relationship between the dark current and the scattering rate, non sharpness of the EBP involves a rise of 20% of the dark current for the typical value of gradient extension obtained via the STEM measurements. In our range of energy and for the operating temperature of our device, phonon processes are inefficient source of transport, see tab I. Clearly, for this structure, the interaction between electron and ionized impurities is the one which drives the transport[6]. The evolution of the intrawell scattering rate for the lowest subband has been simulated, this scattering rate is almost constant to $1.1\times10^{13}$Hz with the gradient extension. Such a result is not really surprising since the overlap of the associated wave function with itself stays equal to one whatever complicated the EBP could be.

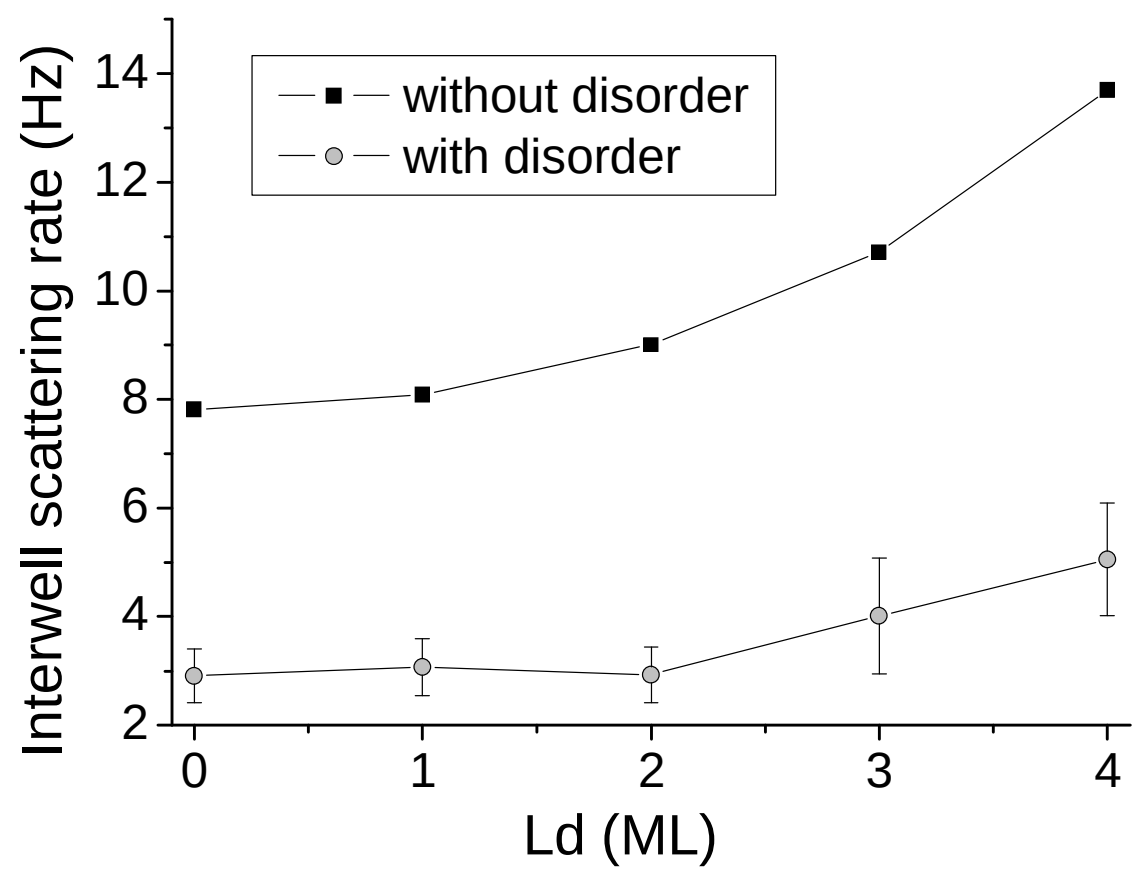

FIG. 5 Interwell scattering rate as a function of the diffusion length with and without disorder at T=10K and under an electric field of 5kV·cm$^{-1}$. For the curve with disorder each point represents the mean value after averaging over fifty charges distribution configurations, the error bar is the associated standard deviation

**E. Influence of the well width and detected wavelength**

The well width dependence of the effect of the non sharpness has to be addressed for the design of robust structure. We should expect that a larger well thickness will be less sensitive to the non sharpness. On the previous structure, we have changed the well width and studied the effect on the energy transition, see FIG. 6. Indeed the graph shows that the larger the well, the lower the effect of the the non-sharpness. On this range of well width the effect seems to follow a linear law.

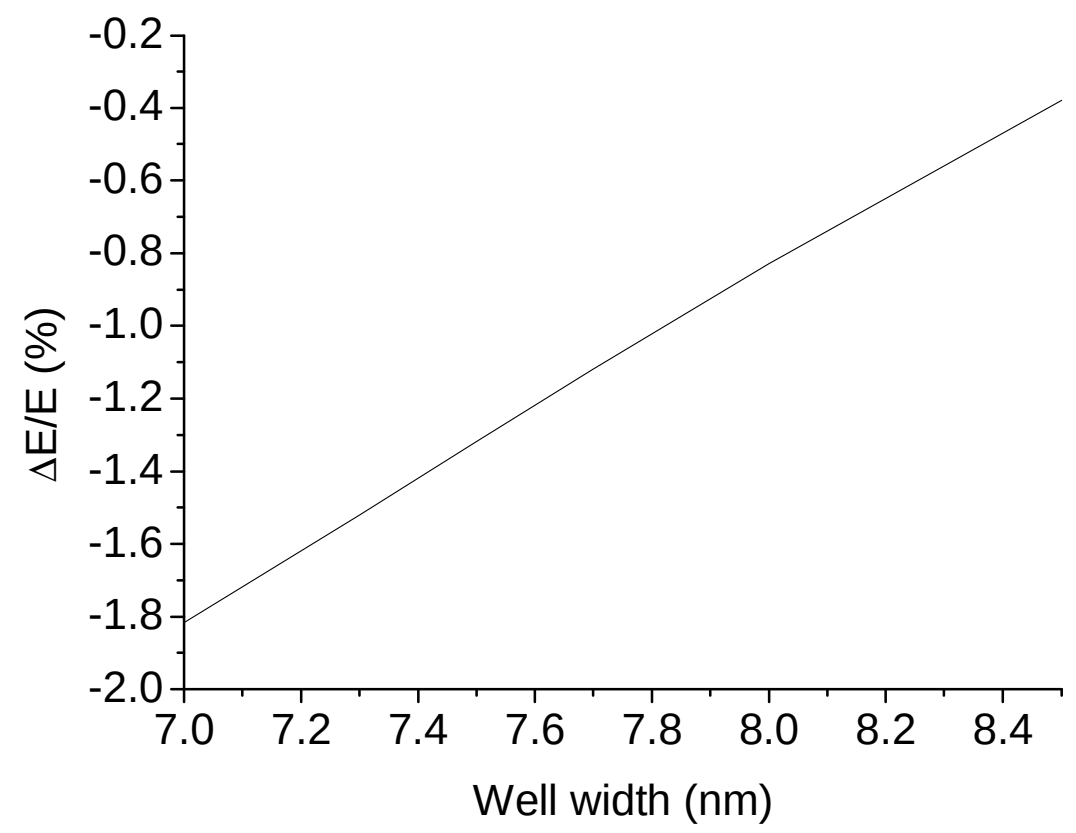

FIG. 6 Difference of the energy transition with and without non-sharpness, normalized by the initial transition energy $\left(\frac{E_{non-sharpness}-E_{ideal}}{E_{ideal}}\right)$, as a function of the well width. The value of the gradient extension is taken equal to 3ML in this graph.

We can generalize our result by studying the effect of the non sharpness for three very different structures given in the literature. The growth parameters are described in the tab. II, they all present a bound to quasi bound transition. The B component is the previously studied component. The FIG. 7 reports the effect of a three monolayers gradient extension on the energy transition. We notice that the higher the detected wavelength, the lower the non sharpness effect. This effect results from the fact that when the electronic confinement in the well is low, the perturbation due to the non sharpness is also reduced.

tab. II Growth parameters of the three components studied in FIG. 7.

| Sample | A | B | C |
|---|---|---|---|
| $L_w$ (nm) | 5 | 7.3 | 11.9 |
| $L_b$ (nm) | 35 | 35 | 55.2 |
| %Al | 26 | 15.2 | 5 |
| Peak wavelength (µm) | 8.5 | 14.5 | 41 |
| Range of detection | LWIR | VLWIR | THz |
| Reference | 16 | | 17 |

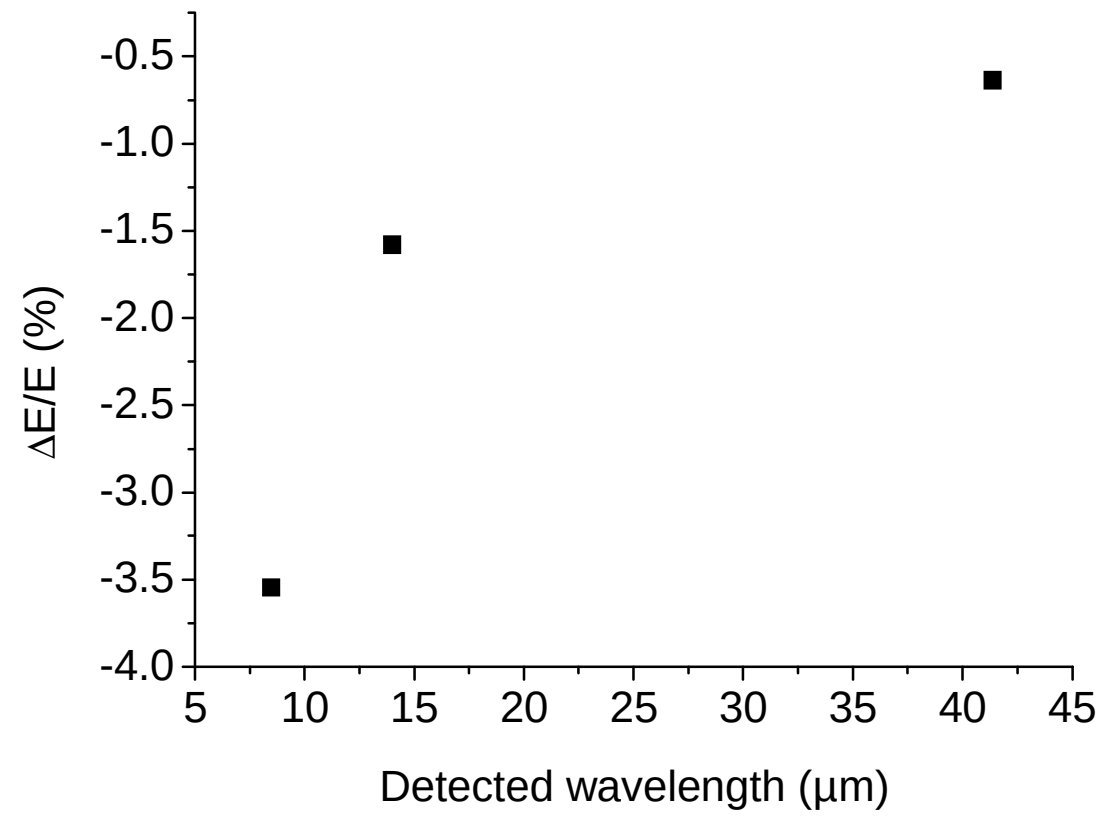

FIG. 7 Difference of the energy transition with and without non-sharpness normalized by the initial transition energy $\left(\frac{E_{non-sharpness}-E_{ideal}}{E_{ideal}}\right)$ for three components described in tab. II. The value of the gradient extension is taken equal to 3ML in this graph.

## IV. Influence of the impurities

In the previous part, we have underlined the role of the non-sharpness of the interface on the EBP. Nevertheless the non steepness of the EBP is not the only source of non ideality. Indeed the presence of impurities also implies a modification of the EBP. Actually the random positioning of the silicon doping in the well plane involves disorder, which leads to a modification of the EBP along the growth direction.

### A. Local energy band profile

To consider the disorder introduced by the ionized impurities on the EBP, the homogenous distribution has to be taken into account as well as its spatial fluctuations. Here we limit the area of the paper to intentionally introduced impurities (doping). We chose to follow a procedure close to the one suggested by Metzner et al.[18] and Willenberg et al[19]. First, we assume that the potential lateral fluctuations are small compared to the variation along the growth direction. Thus the potential linked to these impurities can be split in two parts $V(z,r) = \langle V(z) \rangle + \Delta V(z,r)$ where $\langle V(z) \rangle$ is the biased potential along the growth direction due to the alloy composition and $\Delta V(z,r)$ is the lateral fluctuation due to the disorder. The assumption is being made that we can still write the wave function as the product of an envelope function and a lateral plane wave. Thus rather than evaluating the wave function for different points r, we average the result over a large number of charge distributions, as explained in ref 18.

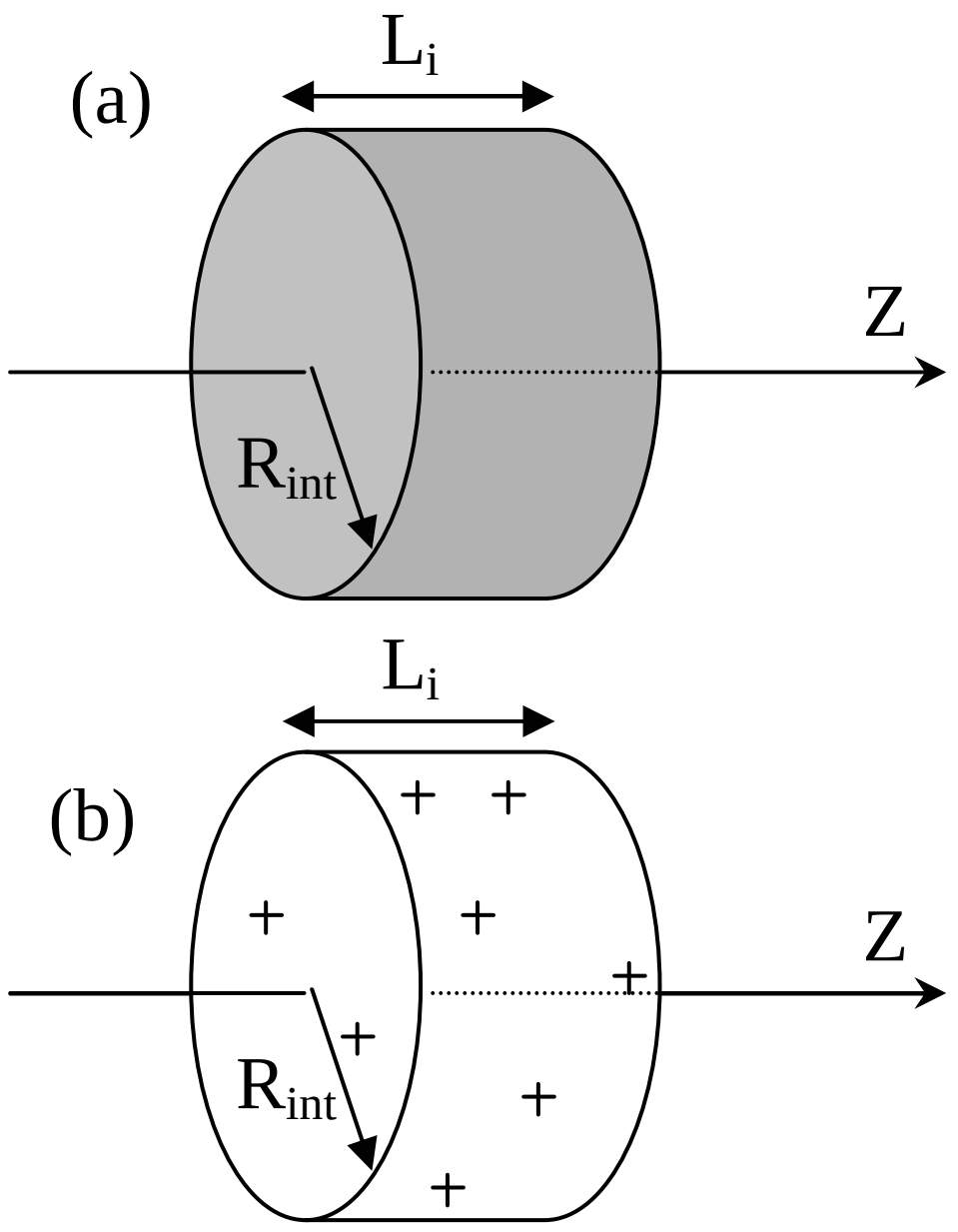

FIG. 8 (a) volume of integration of the screened potential in the case of an homogenous volume distribution, (b) volume of integration of the screened potential in the case of a discrete distribution.

The screened potential is treated as follows. We evaluate the potential along the z axis due to an homogeneously doped cylinder, see FIG. 8 (a). The associated potential is given by[20]:

$$V_{\text{homogenous}}(z) = -\frac{n_{3D} e^2}{4\pi\varepsilon\varepsilon_0} \int_0^{L_i} \int_0^{R_{\text{int}}} 2\pi r' dr' dz' \frac{e^{-q_o r}}{r}$$

where $L_i$ is the doping extension along the growth direction, $r = \sqrt{(z-z')^2 + r'^2}$ and $q_0$ the Thomas-Fermi wave vector given by $q_0{}^2 = \frac{e^2 n_{3D}}{\varepsilon\varepsilon_0 k_b T}$ with $n_{3D}$ the volume density, $\varepsilon$ the relative permittivity, $\varepsilon_0$ the vacuum permittivity, $k_b$ the Boltzmann constant and T the temperature. Here we have considered that the Coulomb interaction is screened in a Thomas-Fermi approach and is thus given by a Yukawa potential. Due to the screening of the potential the radius of the charge distribution stays limited ($R_{int}$ is a few tens of nm)

The potential linked to the discrete charges is given by $V_{\text{inhomogenous}} = -\frac{e^2}{4\pi\varepsilon\varepsilon_0} \sum_{charges} \frac{e^{-q_o r}}{r}$, see FIG. 8(b). We take care to keep unchanged the number of charges considered for the two distributions. Typically the disordered potential becomes independent of the number of charges if N>10 per period.

So the potential linked to the disorder is given by the difference between these two contributions $V_{disorder} = V_{\text{inhomogenous}} - V_{\text{homogenous}}$. As the mean confinement of the electron has to be unchanged, we set the mean value of this distribution equal to zero.[18] The $V_{disorder}$ potential is a step function where the length of each step is equal to one ML. As expected, the disordered potential is important close to the doping region but its extension is much longer than the center of the well, see FIG. 9. A Gaussian fit of the disordered potential distribution leads to standard deviation of 8nm, which is three times longer than the nominal doping region size. Finally to solve the Schrödinger equation this disordered potential is added to the ideal potential.

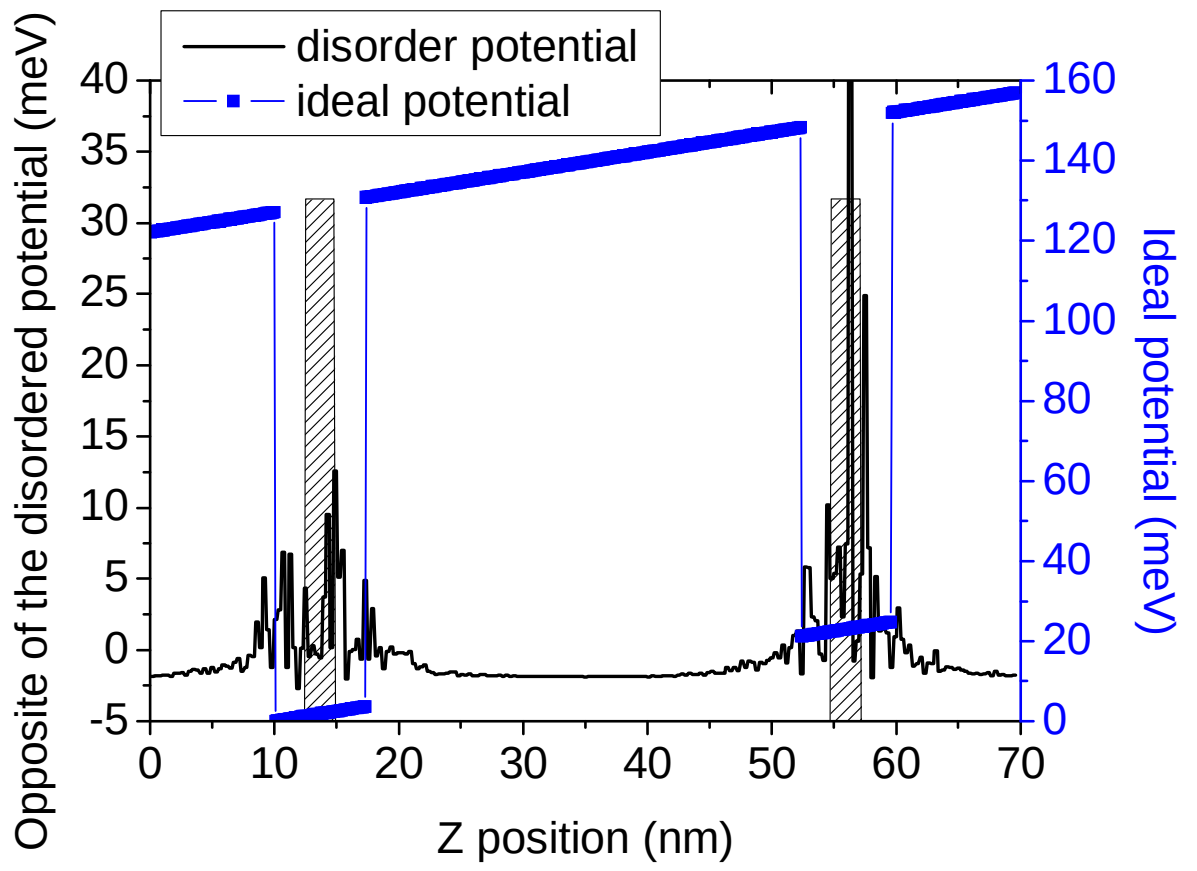

FIG. 9 Opposite of the disordered potential for a given distribution of charge (N=15 charge per well-R~40nm). The striped pattern shows the doping position.

### B. Influence of the disorder on the optoelectronic properties

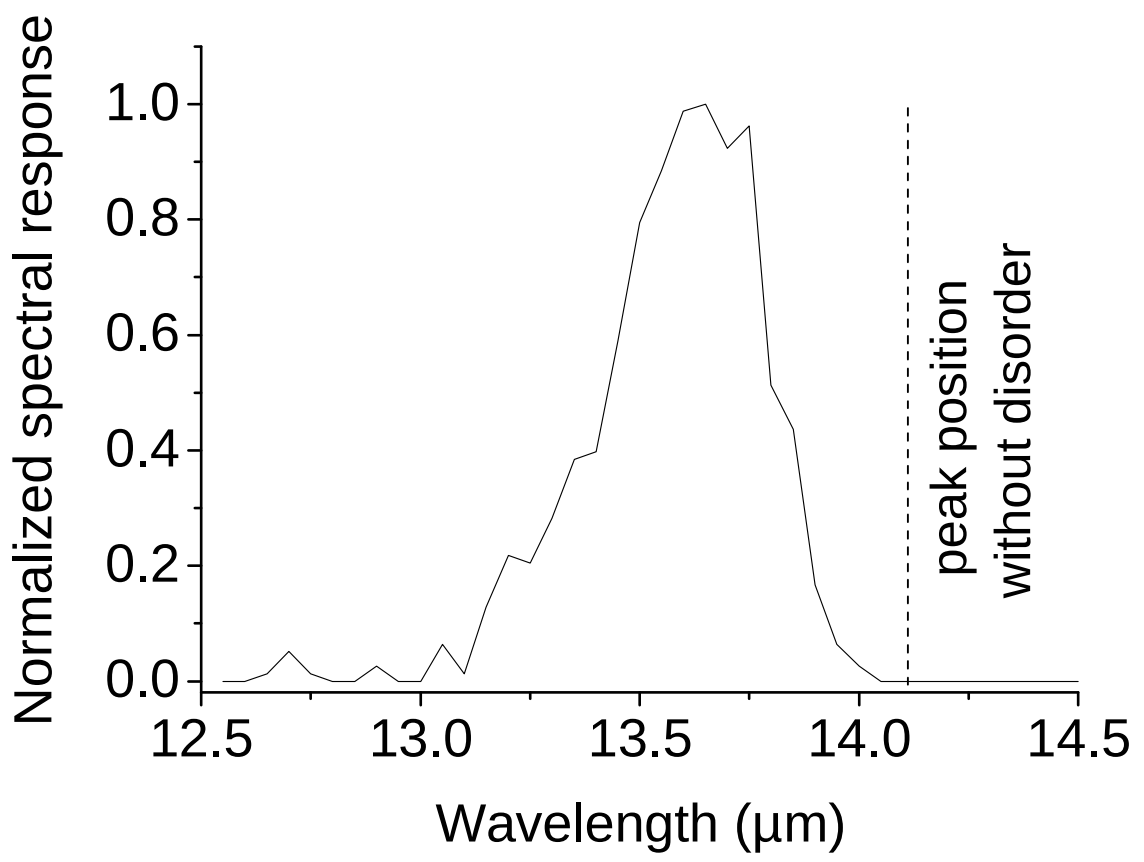

FIG. 10 Spectral response of the QWIP in presence of a inhomogeneous distribution of charges. 700 distributions of charges have been evaluated to obtain this graph.

The presence of donors in the well tends to localize the electron close to the doping (and so into the well) which increases the electron confinement and results in a higher energy difference between the ground state and the first excited state. As a consequence, this disorder leads to a blue shift of the spectral response, see FIG. 10. The spectral response is shifted of -0.5µm. Nevertheless, this blue shift is partly compensated by the red shift due to the roughness of the profile (+0.3µm) making QWIP barely sensitive to the fluctuations of the EBP. The broadening associated with the disorder of the transition is 0.4µm[21] (~3mev), see FIG. 10.

We have also studied the influence of this discreteness of the charges distribution on the interwell scattering rates[22]. Results are presented on FIG. 5 and tab. I. The disorder leads to a reduced dark current, which results from a higher confinement of the electron in the central part of the well. This effect is only slightly counterbalanced by the dark current increase resulting from the EBP non-sharpness. We may also notice that the relative importance of the different scattering process stay almost unchanged.

### C. Importance of the ground state

The ground state energy plays a significant role in the device characteristics. Indeed, the transport properties in the tunnel regime and the energy transition are driven by its energy position.

Since the associated wave function is confined close to the doping and its disorder, the ground state energy will be more sensitive to the bottom of the EBP than the excited state could be.

FIG. 11 shows the energy distribution of each subband in presence of disorder. The ground subband is more affected by the disorder than the excited subband. The shift of the energy transition as well as the broadening of this transition mostly reflects the effect of the disorder over the ground subband.

As a consequence, the design of QWIP dedicated to operate in their tunnel regime has to pay attention on the position of this subband.

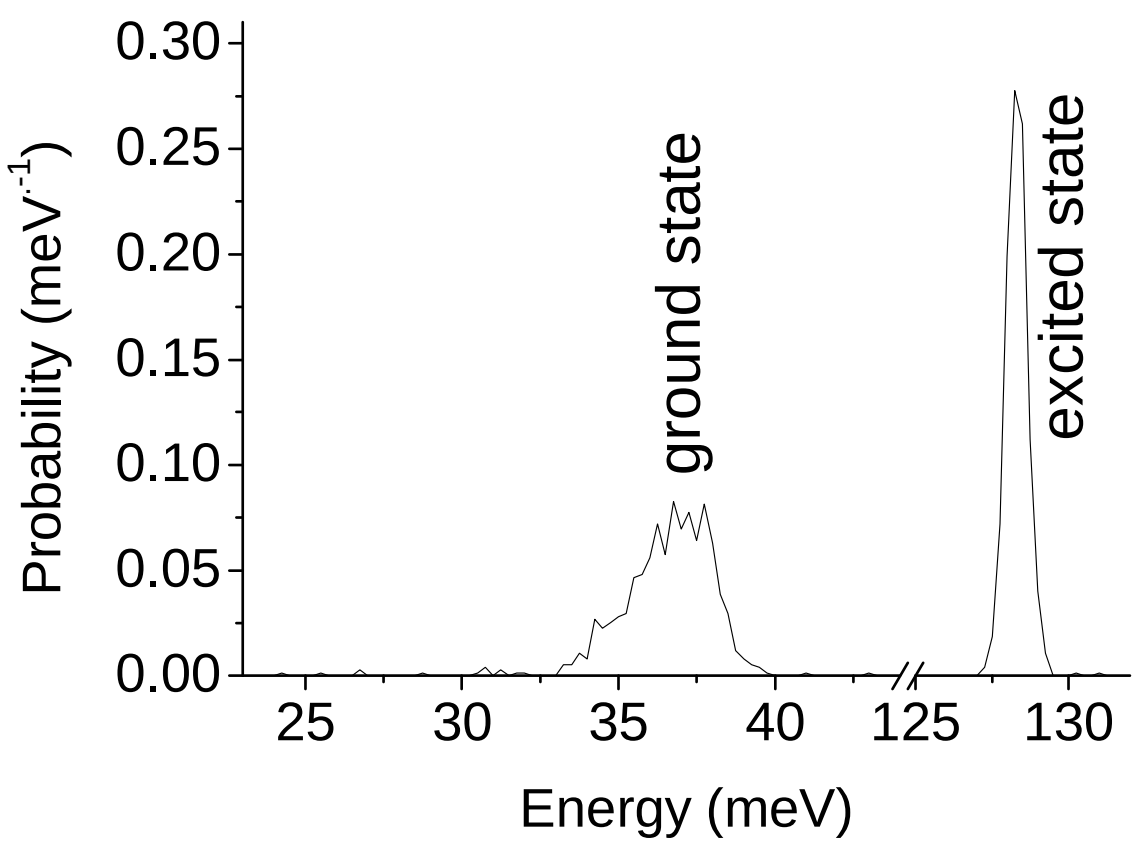

FIG. 11 Probability to find the ground (excited) subband as a function of the energy. 700 distributions of charge have been evaluated to obtain this graph.

## V. Other possible mechanisms of broadening

We notice that the non-sharpness of the EBP (broadening below 1mev) and the disorder due to the presence of doping (broadening around 3mev) are not sufficient to explain the experimental broadening of the transition (around 10meV). Other mechanisms are consequently involved on the broadening of the transition.

Another possibility of non ideality which may result in a broadened transition is the roughness of the interface[23]. We do not expect that this roughness involved a shift of the detected wavelength since the in plane dependence of the interface coordinate presents a null mean value. Interface roughness parameters have been measured in a previous work[24], the magnitude of the roughness is in the two-three monolayers range whereas the mean distance between defects is around 10nm. This results in 2.4meV broadening, see tab. III, which makes interface roughness one of the main processes for the broadening of the transition.

It is generally admitted that the electric field distribution in QWIP is not perfectly homogenous over the whole structure and changes of 20% of the value of the electric field have already been reported[25]. This may imply a change in an inhomogeneous broadening of the transition which has been evaluated to 0.1meV, see tab. III.

We have also evaluated the scattering rate associated to this transition for LO phonon, LA phonon, alloy disorder and an homogeneous distribution of charges, see tab. III. Their effects stay moderate. The inter electrons broadening has been considered using the procedure described in reference 12 and 26. This process clearly appears has the dominant interaction for the broadening of the transition. But such a model is known to be optimistic.

To finish we have investigated the effect of the non parabolicity[27]. This effect involved a shift of the transition (redshift of 1.2 meV) and 0.2meV broadening of the transition.

Finally the sum of the different broadening processes results in a 14.2 meV broadening which is quite close to the experimental value (12 meV).

tab. III Broadening of the transition associated with different processes

| Process | Broadenning (meV) |
|---|---|
| LA phonon | $1.1\times10^{-2}$ |
| Ionized impurities | 0.1 |
| Electric field | 0.1 |
| Non parabolicity | 0.2 |
| Alloy disorder | 0.9 |
| Non-sharpness | 0.9 |
| LO phonon | 1.6 |
| Interface roughness | 2.4 |
| Charge disorder | 3 |
| Inter-electron | 5 |

## VI. Conclusion

We have investigated two different sources of non ideality in QWIP structure, which are the non-sharpness of the alloy profile and the disorder induced by the discrete character of the charges. STEM measurements were used to measure the smoothing length, which is evaluated to two or three monolayers. The smoothing tends to reduce the confinement of the electron in the well and finally leads to a redshift of the spectral response and an increase of the dark current. At the opposite the discreteness of the charges distribution increases the electron confinement near the doping and consequently into the well in the case of QWIP. Thus, this disorder involves a blue shift of the spectral response and reduces interwell scattering rate. Finally, those two opposite effects partly compensate each other, which makes QWIP only poorly sensitive to the details of the EBP. We thus showed that it is not necessary to include those effects for a first order evaluation of the QWIP properties. The use of a non ideal EBP is only required if the expected accuracy on the QWIP properties is better than a few percent.